\documentstyle[12pt,aps,epsf]{revtex}
\begin{document}
\baselineskip=17pt
\draft
\title{Prospects for observations of high-energy cosmic tau neutrinos}
\author{H. Athar\footnote{Present Address: Department of Physics, 
        Tokyo Metropolitan University, 1-1 Minami-Osawa, Hachioji, 
        Tokyo 192-0397, Japan.}, G. Parente and E. Zas}\address{ 
        Departamento de F{\'\i}sica de Part{\'\i}culas, 
 Universidade de Santiago de Compostela, E-15706 Santiago de Compostela,
 Spain; E-mail: athar@fpaxp1.usc.es, gonzalo@fpaxp1.usc.es, zas@fpaxp1.usc.es}
\date{\today}
\maketitle
\begin{abstract}
\tightenlines

 We study  prospects for the observations of high-energy cosmic 
tau neutrinos ($E\, \geq 10^{6}$ GeV) originating from proton acceleration in 
the cores of active galactic nuclei. 
We consider the possibility that vacuum flavor neutrino 
oscillations induce a tau to muon neutrino flux ratio greatly exceeding 
the rather small value expected from intrinsic production. 
 The criterias and event rates 
for under water/ice light \v{C}erenkov neutrino telescopes are given by 
considering the possible detection of downgoing high-energy cosmic tau 
neutrinos through characteristic double shower events. 
\end{abstract}


\section{Introduction}

Neutrino astronomy is now an emerging field and 
entails the need to have improved flux estimates 
as well as a good understanding of relevant
detector capabilities for all flavor neutrinos, particularly 
in light of the recent growing experimental support 
for flavor oscillations \cite{YU}. 
 Several high-energy cosmic neutrino ($> \sim 10^6~$GeV) 
detectors based on under 
water/ice muon detection are now at proposal or construction 
stages \cite{TASI} and alternative techniques for $10^9$~GeV 
neutrinos are being considered
through coherent radio \cite{Zal} or acoustic \cite{zelez} 
pulses as well as through horizontal air shower measurements 
with conventional arrays \cite{Capelle} or with 
fluorescent light either from the ground \cite{Hires,TA} 
or from orbiting detectors \cite{orb}. A number of astrophysical 
high-energy neutrino sources, such as Active Galactic Nuclei (AGN), 
have been discussed in the literature and predicted to produce fluxes 
that could be detected in some of these detectors \cite{physrep}. 
In the event of a successful detection of high-energy 
astrophysical neutrinos the range of parameter space for 
flavor oscillations that can be tested could be considerably 
enhanced provided a flavor identification can be done. 

Cosmic tau neutrinos are possibly the easiest flavor to identify 
above 10$^6$ GeV. Two ideas have already been put forward 
based on the short decay lifetime of the $\tau$ 
produced in charged current interactions. 
There is a suggestion of measuring 10$^6$ GeV 
$\nu_{\tau}$ flux through double shower ({\sl double bang}) events 
 in under water/ice \v Cerenkov telescopes \cite{L}. 
A more recent suggestion is to detect a small 
pile up of upgoing $\mu$-like events in the 10$^{4}-$ 10$^{5}$~GeV 
range with a fairly flat zenith angle dependence \cite{HS}. 
On the other hand the LPM effect, that lengthens electromagnetic showers 
in water and ice, has been suggested to separate 
electron neutrino charged current interactions from the rest. 
This could be done with \v{C}erenkov light detectors 
or with the radio technique  for energies 
above $2\cdot 10^{7}$ GeV \cite{Zal99}. 
It is conceivable that a combination of several of 
such techniques will allow the establishment of 
neutrino flavor ratios at energies above $10^6$~GeV.

We will concentrate on cosmic tau neutrino detection in this article. 
Specifically, we discuss in some detail, the prospects for detection 
of high-energy cosmic tau neutrinos originating from the cores of AGN. 
We consider vacuum flavor oscillations as an example to illustrate 
the possibility offered by detectors in construction to distinguish 
between different neutrino flavors. 
For absolute event rates we use upper limit flux calculations as 
an example \cite{szabo} and consider $1~$km$^2$ detector 
sizes which are now being planned \cite{TASI}. 
Both the energy ranges of interest and the 
relative numbers of $\tau$- and $\mu$-like event rates are 
however independent of the assumed normalization of the neutrino fluxes.

We show that for the chosen neutrino flux, 
a km$^{2}$ size surface area under ice/water \v{C}erenkov
light neutrino detector may be able to either set useful upper 
limits or may obtain first examples of the high-energy tau neutrinos 
originating from this cosmologically distant astrophysical source. 
The plan of the paper is as follows: In Section II, after a brief 
discussion of intrinsic production mechanisms of high-energy muon and 
tau neutrinos in AGNs, we estimate the relevant vacuum flavor oscillation 
 probability. In Section III, we discuss in some detail the
detection technique making use of the double shower structure of the 
tau neutrino charged current interactions and calculate the 
expected event rates for 
a typical km$^2$ surface size under water/ice detector. 
In Section IV, we summarize our results. 

\section{Intrinsic Cosmic Tau Neutrino Production and vacuum flavor 
         oscillations}

AGNs are the brightest objects in the sky and high-energy photons 
reaching tens of thousands of GeV have been observed from them. 
This is commonly interpreted as an indication that some kind of 
Fermi acceleration is taking place. In conventional models, electrons 
are the particles that get accelerated. 
It has been argued that if Fermi mechanisms are able to accelerate 
electrons in these objects, protons could also be accelerated by them. 
In proton acceleration models, the photons arise from neutral pion decays 
either in $p \gamma$ or $pp$ collisions. 
If this is true, electron and muon neutrinos are also expected to be 
produced at similar flux levels from charged pion decays. Neutrino 
detection can provide the signature for proton acceleration in AGN 
and it is one of the main goals of neutrino detectors in 
construction and design stages. 
For an update review of various $\nu_{e}$ and $\nu_{\mu}$ 
flux estimates from AGNs in $p\gamma $ and $pp$ collisions  
(as well as from some other interesting astrophysical/cosmological sites), 
see \cite{nu98}. 
We will here explore the expected tau neutrino fluxes intrinsically 
produced in the collisions and how these fluxes can vary in a possible 
neutrino oscillation scenario. 

In $p\gamma $ collisions,
high-energy $\nu_{e}$ and $\nu_{\mu}$ are mainly produced through the 
resonant reaction $p+\gamma \rightarrow \Delta^{+}\rightarrow n+\pi^{+}$. 
The same collisions will give rise to a greatly suppressed 
high-energy $\nu_{\tau}$ (and $\bar{\nu}_{\tau}$) flux mainly through the 
reaction $p+\gamma \, \rightarrow \, D^{+}_{S}+\Lambda^{0}+\bar{D}^{0}$. 
The production cross-section for $D^{+}_{S}$ is essentially up to three
orders of magnitude lower than that of $\Delta^{+}$ production 
for the relevant center of mass energy scale. 
Moreover the branching ratio of $D^{\pm}_{S}$ to decay eventually into 
$\nu_{\tau}$ ($\bar{\nu}_{\tau}$) is approximately two orders of magnitude 
lower than for $\Delta^{+}$ to subsequently decay into $\nu_{e}$ and 
$\nu_{\mu}$ through $\pi^{+}$. These 
two suppression factors along with the relevant kinematic limits give 
approximately the ratio of intrinsic fluxes of tau neutrinos and 
muon neutrinos as: $(\nu_{\tau}+\bar{\nu}_{\tau})/
 (\nu_{\mu}+\bar{\nu}_{\mu})\, <\, 10^{-5}$.

In $pp$ collisions, the $\nu_{\tau}$ flux may be obtained through 
$p+p\rightarrow D^{+}_{S}+X$.
The relatively small cross-section for $D^{+}_{S}$ production 
together with the low branching 
ratio into $\nu_{\tau}$ implies that the 
$\nu_{\tau}$ flux in $pp$ collisions is also suppressed up to $4-5$ 
orders of magnitude relative to $\nu_{e}$ and/or $\nu_{\mu}$ fluxes. 
The situation is quite similar to the prompt atmospheric 
$\nu_{\tau}$ flux calculation; the result is basically a rescaling of the 
prompt $\nu_{\mu}$ flux from the decay of charmed $D$'s and results in a 
negligibly small  $\nu_{\tau}$ flux for the energies under discussion 
\cite{pasqua,con}. 

In proton acceleration models the intrinsically produced 
tau neutrino flux is thus expected to be very small, typically 
a factor between $10^{-5}$ and $10^{-6}$ relative to electron 
and muon neutrino fluxes \cite{vaz}.
However, recent experimental measurements of atmospheric neutrinos 
suggest that neutrinos could just have vacuum flavor oscillations and 
the tau neutrino flux would be dramatically enhanced. 

It has been pointed out that there are no matter effects for high-energy 
cosmic tau neutrinos originating from cores of AGNs primarily because of 
relatively small matter density in the vicinity of core of the AGN 
 for all relevant $\delta m^{2}$ \cite{japan}. 
We will restrict the following discussion to vacuum oscillations between 
two flavors, $\nu_{\mu}$ and $\nu_{\tau}$, for simplicity. 

The flavor precession probability 
for non vanishing vacuum mixing angle is obtained from the 
effective Hamiltonian matrix in the two flavor basis 
$\psi^{T}\, =\, (\nu_{\mu}, \nu_{\tau})$:
\begin{equation}
\left( \begin{array}{cc}
    0                 & \frac{\delta}{2} \sin2\theta    \\
    \frac{\delta}{2} \sin2\theta   & \delta \cos2\theta
    \end{array}\right),
\label{Htau}
\end{equation}
where $\delta  = \delta m^{2}/{2E}$ with 
$\delta m^{2} = m^{2}(\nu_{\tau})-m^{2}(\nu_{\mu})$
and $E$ the neutrino energy, leading to the well known result:
\begin{equation}
  P(\nu_{\mu}\rightarrow \nu_{\tau})\, =\, \sin^{2}2\theta 
 \sin^{2}\left(\frac{\delta m^{2}}{4E}L\right).
\end{equation}
If we take the values of $\sin^{2}2\theta $ and $\delta m^{2}$ suggested by 
recent superkamiokande data ($\sin^{2}2\theta \, \sim 1, \delta m^{2}\, 
 \sim \, 10^{-3}$ eV$^{2}$) \cite{skk} and  $L\, \sim \, $
100 Mpc (1~pc $\simeq 3\cdot 10^{18}$ cm) as a representative 
distance between the AGN and our galaxy, then the above rapidly 
oscillating probability averages out to $\, \sim \, 1/2$ for all relevant 
neutrino energies to be considered for detection. Very similar 
fluxes of muon and tau neutrinos would thus be expected.   
Let us further note that after averaging the $P$ given by Eq. (2) is 
independent of not only $E$ but also $\delta m^{2}$ and thus leads to a 
constant suppression of high-energy cosmic muon neutrino flux.
 
The deficit measured by 
superkamiokande in atmospheric  muon
 neutrino flux may currently be explained either through 
 $\nu_{\mu}\rightarrow \nu_{\tau}$ or through $\nu_{\mu}\rightarrow 
 \nu_{s}$, where $\nu_{s}$ is a sterile neutrino\footnote{Although, $\nu_{\mu}
 \rightarrow \nu_{s}$ is now being disfavoured \cite{nakahata}.}. 
In the first case and
for high-energy  neutrinos originating at cosmological distances,
the ratio $(\nu_{\tau}+\bar{\nu}_{\tau})/
 (\nu_{\mu}+\bar{\nu}_{\mu})\, $ is close to 1/2. 
Therefore, a ratio different from 1/2 excludes this possibility. 
  
\section{Detection of high-energy cosmic tau neutrinos}

High-energy cosmic tau neutrino detection could be achieved by  
making use of the characteristic double shower events \cite{L} 
or by the pileup effect expected as they travel through 
the Earth \cite{HS}. 
Such events could be seen in conventional neutrino 
telescopes and in principle also with 
other alternative techniques that have been proposed. 
We will discuss in some detail the possibility 
of detecting double shower events for conventional 
underground telescopes by estimating rates using the the 
fluxes of Ref. \cite{szabo} and the oscillation probability 
addressed in the previous section. This is intended 
to provide a reference calculation.

The downgoing cosmic tau neutrinos reaching close to the 
surface of the detector may undergo  a charged current 
 deep inelastic 
scattering with nuclei inside/near the detector and produce 
 a tau lepton in addition
to a hadronic shower. This tau lepton
traverses a distance, on average proportional to its energy, 
before it decays back into a tau 
neutrino and a second shower most often induced by decaying  hadrons. 
The second shower is expected to carry about twice as much energy 
as the first and such double shower signals are commonly referred 
to as double bangs. 
As tau leptons are not expected to have further relevant 
interactions (with high-energy loss) in their decay 
timescale, the two showers should be 
 separated by a {\sl clean} $\mu$-like track \cite{L}. 

We are going to restrict our estimate to downgoing neutrinos as at these 
energies tau neutrinos that go through the Earth interact. Effectively 
the process of interaction and tau decays can be regarded as an energy 
degradation to the range $10^4-10^5$~GeV \cite{HS}. Unfortunately the 
two shower signature will be difficult to be resolved below 
$\sim 3\cdot 10^5~$GeV (see below). 

For downgoing cosmic tau neutrinos, the {\em double bang} event rate in 
water/ice is  estimated using \cite{gq}
\begin{equation}
 {\mbox{Rate}}=A\int
 \mbox{d}EP_{\tau}(E,E^{{\rm{min}}}_{\tau})\frac{\mbox{d}N}{\mbox{d}E},
\end{equation}
where $A$ is the area of the neutrino telescope 
and $P_{\tau}$ gives the 
probability that a tau neutrino of energy $E$ produces two contained and
separable showers with the tau lepton energy
greater than $ E^{{\rm {min}}}_{\tau}$.
It is given by:
\begin{equation}
 P_{\tau}(E,E^{{\rm {min}}}_{\tau}) = \rho_{\rm {m}} \cdot N_{A}
\int^{1-{E^{{\rm {min}}}_{\tau} \over E}}_{0}\mbox{d}y
 \left[D- R_{\tau}\right]
\frac{\mbox{d}\sigma^{CC}(E,y)}{\mbox{d}y},
\end{equation}
where $N_{A}$ is the Avogadro's number, $\rho_{\rm {m}}$ is the density of
the detector medium,
$\mbox{d}\sigma^{CC}/\mbox{d}y$ is the charged current $\nu_{\tau}N$ 
differential cross-section,
$y$ is the fraction of neutrino energy that is transferred to the 
hadron in the laboratory frame. $D$ is the detector length scale which 
we fix to be $1~$km and the 
tau lepton range $R_{\tau}$ which must be contained within $D$ is given by:
\begin{equation}
R_{\tau} = \frac{E(1-y)\tau{\mbox{c}}}{m_{\tau}{\mbox c}^{2}}.
\end{equation}
In Eq. (5), $\tau {\mbox c}$ is the lifetime and 
 $m_{\tau}{\mbox c}^{2}$ is the mass of the high-energy tau lepton.

The lower limit of integration in Eq. (3) is $E^{{\rm{min}}}_{\tau}$.
We take $ E^{{\rm{min}}}_{\tau}$ greater than $\sim \, 2\cdot 10^{6}$ GeV
because at this energy
the tau lepton range (separation between the two showers) 
in water is $\sim $ 100 m which allows a clear separation between 
the two showers. The upper limit of integration in Eq. (3) 
is taken to be $E\,{\buildrel <\over {_{\sim }}} \, 2\cdot 10^{7}$~GeV
as for energies above it the tau lepton range exceeds the telescope 
size (see  Fig. 1).

Finally in Eq. (3), d$N/$d$E$ is the differential high-energy
tau neutrino flux and is obtained by multiplying   
 $P(\nu_{\mu}\rightarrow \nu_{\tau})$ given by Eq. (2) with d$N$/d$E$ for 
 $(\nu_{\mu}+\bar{\nu}_{\mu})$ taken from \cite{szabo}.
 In Fig. 2, we depict downgoing differential event rates for 
 double shower events using the  
 parton distributions MRS R$_{1}$ from \cite{mrs} for km$^{2}$ 
 under water neutrino telescopes as an example.
 We have checked that other modern parton distributions give quite 
 similar events rates and are therefore not depicted here. 
We have taken into account the fact that $\sim 15 \% $ of the times the
tau decay does not induce any shower. 
For comparison we also plot the $\mu $-like event rate induced by 
muon neutrinos. Note that these $15\%$ and the 
 tau neutrino interactions in which the tau lepton 
decays outside the detector volume have identical ($\mu $-like) experimental 
signature.

The signature of double shower events depends on the detector capabilities 
for shower identification and energy resolution and difficulties can be 
envisaged. We have used 100~m as the minimum distance to resolve 
two showers, what is quite conservative in view of typical spacing between 
Optical Modules in an under water/ice detector. 
In Fig. 3, we show the dependence of shower size and shower separation
on neutrino energy $E$ in ice and in water for which we have used the 
parametrization of \cite{Gaisser}. 
As shower size is basically proportional to energy, the size of the
second shower is on average a factor of 2 
higher than the first one (see also \cite{L}). This
value results by taking into account the relevant kinematics of
the allowed decay channels and the corresponding branching ratios and 
using the average energy transfer $\langle y\rangle =0.25$. The 
$y$ distribution and decay kinematics will lead to a spread in this 
ratio. While $y=0.1$ enhances the energy ratio of the second and first 
showers to a value of about 6, for $y$ values higher than 0.4 the ratio of 
the two shower sizes starts to be lower than unity obscuring the tau 
neutrino signature. 

Another relevant point is the evaluation of the backgrounds, 
a double shower signature not induced by a tau neutrino.
As it was discussed in  \cite{L} such probability is very small and
should not affect the detection of the high-energy cosmic tau neutrino. 
Also one should take into account the possibility that the muon component 
 of a single
cascade induced from a muon or an electron neutrino charged/neutral 
 current interaction can be confused with
the second shower of the tau lepton decay. However, in this case
the size of the second shower is smaller than the first one which 
should be sufficient to distinguish it from a tau neutrino event.
Thus, the selection criteria of amplitude of second shower greater than the
first one typically by a factor of $\sim $ 2, depending on $y$ value,
essentially makes the observation background free.   
 
The high-energy neutrino telescopes have quite small double shower event
 rates (yr$^{-1}$sr$^{-1}$) due to small high-energy intrinsic tau 
 neutrinos flux, thus any observed change in this situation may provide
 indirect evidence of neutrino mass. 
 A corresponding comparable change in this situation is currently not expected
 from variations of astrophysical model inputs.  
The almost simultaneous measurement
 of the two showers may provide useful information on the incident neutrino 
energy as well as the $y$ distribution.

Summarizing, in the context of relevant backgrounds, we envisage 
 essentially the simultaneous presence of two types of events (with different
 topologies)  
serving as background for the tau neutrino induced contained but separable 
 double showers connected by a $\mu$-like track such that the amplitude 
of the second shower is typically 2 times the first one. The first type of 
background events are due to relatively long ($\sim $ 10 km) range muons 
 passing 
 through the detector identified as $\mu$-like tracks. Their estimated 
number is given by the upper slanted curve in Fig. 2. The second type of
background is the single showers due to charged/neutral current interactions.
These may be estimated as 1/10 of the continuous $\mu$-like tracks. 
Thus, the signature of the tau neutrinos as emphasized earlier remained 
distinct from these two type of backgrounds.

We emphasize that within the respective energy window, the essential 
 factor in 
prospective detection of contained but separable double shower events 
  connected by a 
$\mu$-like track as a signature of tau neutrinos is the {\em difference} in the
incident tau neutrino energy dependences on spread and separation of the 
 two showers. 
This {\em difference} is also clearly crucial for separating the tau neutrino 
 events from 
the (relatively abundent) $\mu$-like events.

\section{Results}

	 The intrinsic fluxes of the high-energy cosmic neutrinos 
originating from proton acceleration in cores of AGNs   are estimated
to have typically the following ratios: 
$ (\nu_{\tau}+\bar{\nu}_{\tau})/
 (\nu_{\mu}+\bar{\nu}_{\mu})\, < \, 10^{-5}$. 
 Thus, if an enhanced $(\nu_{\tau}+\bar{\nu}_{\tau})/
 (\nu_{\mu}+\bar{\nu}_{\mu})$
 ratio (as compared to no precession situation) is 
observed {\em correlated} to the direction
of source for high-energy cosmic  
neutrinos, then it may be an evidence for 
 vacuum flavor oscillations of neutrinos induced
 by non zero vacuum mixing angle depending on the finer 
 details of the relevant high-energy cosmic neutrino spectra.
 For vacuum flavor
 oscillations of high-energy cosmic neutrinos, the relevant range of neutrino 
mixing parameters are:  $\delta m^{2}\, \sim 
\, 10^{-3}$ eV$^{2}$ with $\sin^{2}2\theta \, \sim \, 1$.

We have identified the incident tau 
neutrino energy range and the relevant neutrino mixing parameters which may 
give rise to high-energy cosmic tau neutrino induced downward contained but 
separable  double shower events. 
 For  $2\cdot 10^{6}\, {\buildrel < \over {_{\sim}}}\, E/\mbox{GeV}\,
{\buildrel < \over {_{\sim}}} \, 2\cdot 10^{7}$, a km$^{2}$ detector may  
 be able to obtain first examples of downgoing high-energy cosmic tau
 neutrinos through contained but separable double shower events or may at 
 least provide some useful  relevant upper limits.

\paragraph*{Acknowledgments.}

The authors acknowledge the financial support from 
Xunta de Galicia (XUGA-20602B98) and CICYT (AEN96-1773).  
H. A. also thanks Agencia Espa\~nola de Cooperaci\'on Internacional 
(AECI) and Japan Society for the Promotion of Science (JSPS) for financial 
support.

\pagebreak

\begin{figure}[t]
\leavevmode
\epsfxsize=3.5in
\epsfysize=3.5in
\epsfbox{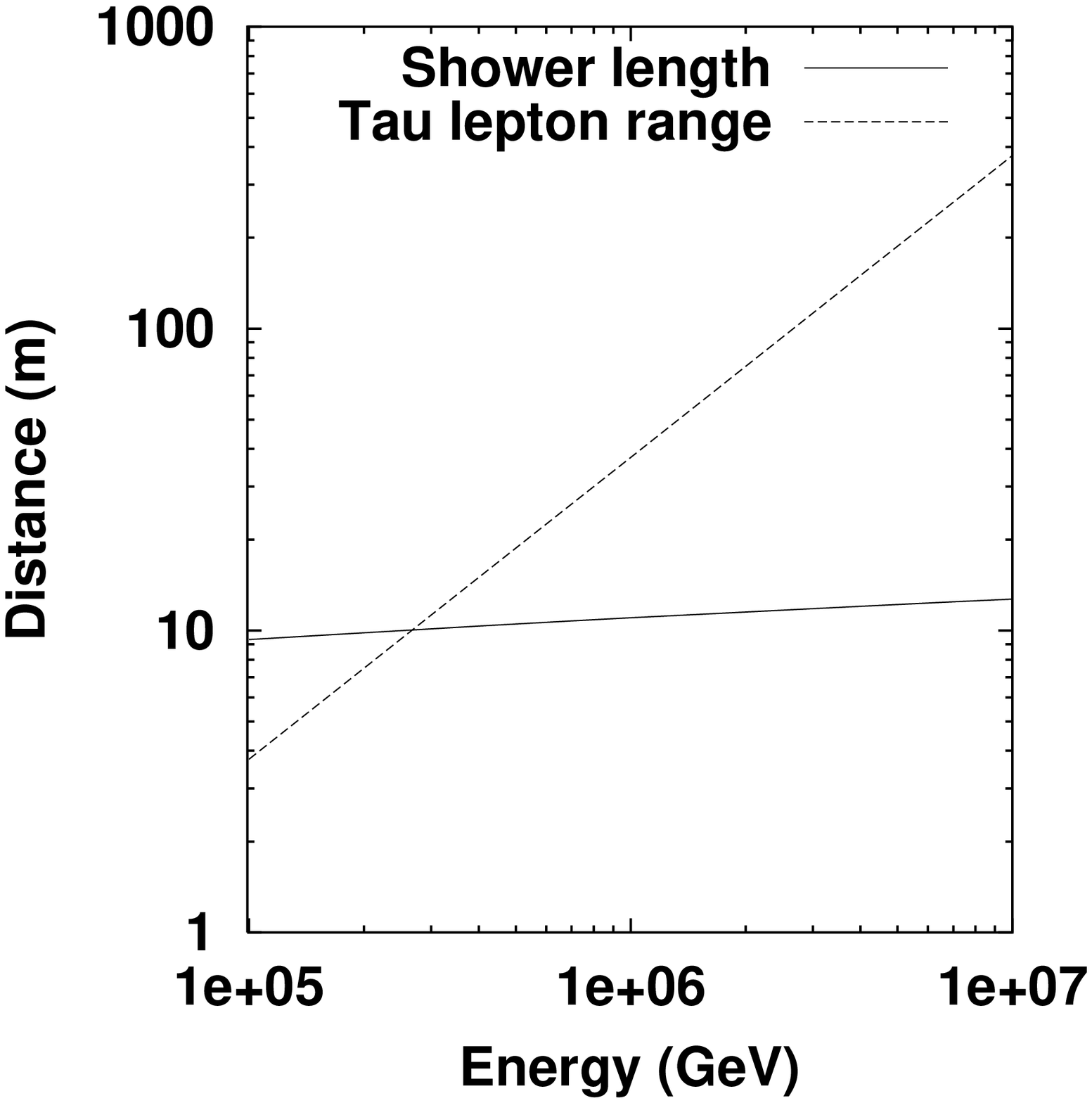}
\tightenlines
\caption{Comparison of the tau lepton range and the shower length of the
 first shower (defined as twice the depth at maximum) for ice/water.}
\end{figure}

\begin{figure}[t]
\leavevmode
\epsfxsize=3.5in
\epsfysize=3.5in
\epsfbox{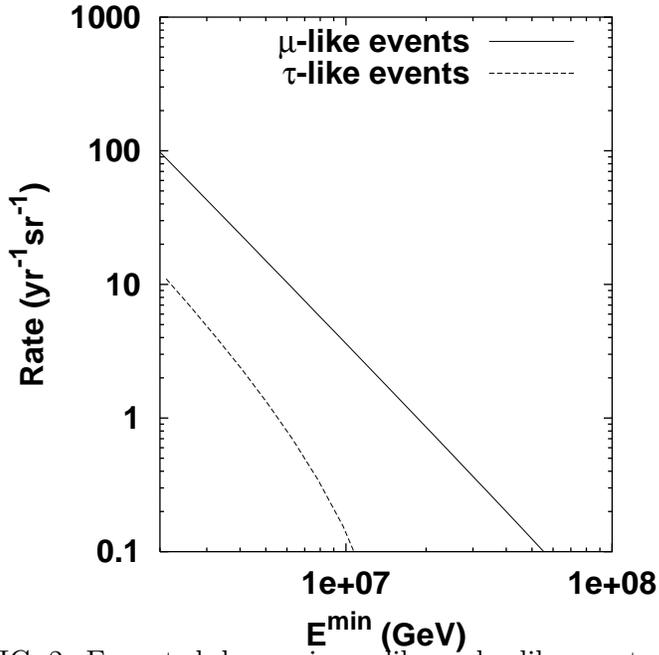}
\tightenlines
\caption{Expected downgoing $\mu$-like and $\tau$-like event rate produced by
 AGN neutrinos from ref. [13] because of vacuum flavor oscillations 
 as a function of the minimum energy of the corresponding lepton.}
\end{figure}

\pagebreak

\begin{figure}[t]
\leavevmode
\epsfxsize=3.5in
\epsfysize=3.5in
\epsfbox{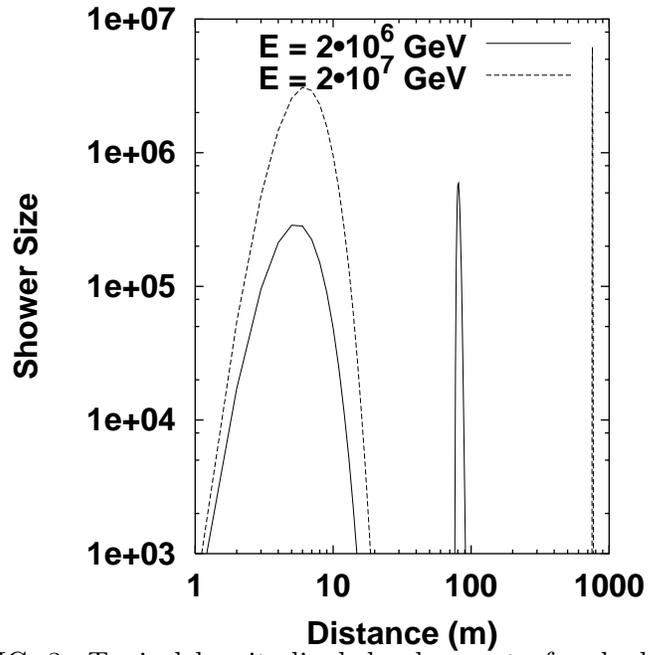}
\tightenlines
\caption{Typical longitudinal development of a double  shower 
 produced by the deep inelastic charged current interaction of the 
 high-energy cosmic tau neutrino.}
\end{figure}

\end{document}